# THERMODYNAMICS OF THE INTERACTING FERMI-SYSTEM IN THE STATIC FLUCTUATION APPROXIMATION

Nigmatullin R.R., Khamzin A.A. and Popov I.I.

Theoretical Physics Department, Kazan (Volga Region) Federal University, 420008, Kremlevskaya str.18

Kazan, Russian Federation. E-mail: nigmat@knet.ru; ii.popov@bk.ru

**Abstract**

We suggest a new method of calculation of the equilibrium correlation functions of an arbitrary order for the interacting Fermi-gas model in the frame of the static fluctuation approximation (SFA) method. This method based only on the single and controllable approximation allows to obtain the so-called far-distance equations (FDEs). These equations connecting the quantum states of a Fermi particle with variables of the local field operator contains all necessary information related to calculation of the desired correlation functions and basic thermodynamic parameters of the many-body system considered. The basic expressions for the mean energy and heat capacity for electron gas at low temperatures in the limit of high density were obtained. All expressions are given in the units of $r_s$, where $r_s$ determines the ratio of a mean distance between electrons to the Bohr radius $a_0$. In these expressions we calculated the terms of the order $r_s$ and $r_s^2$, correspondingly. It was shown also that the SFA allows to find the terms related with high orders of the decomposition with respect to the parameter $r_s$.

---

[1] Corresponding author, E-mail: nigmat@knet.ru

# 1. INTRODUCTION TO THE SFA METHOD

The calculation of the correlation energy of the ground state in the absence of external fields for the strongly degenerated electron gas constitutes the central problem of the solid state physics, in particular the physics of metals. In the pioneering works of Gell-Mann, Brueckner [1] and Wigner [2] the first analytical results for high and low density electron gas were obtained. In these papers an expression for the ground state energy has been obtained. This expression was expressed in the terms of the dimensionless parameter $r_s = r_0 / a_0$, where $r_0 = \sqrt[3]{3/(4\pi n)}$ is the mean distance between electrons, $n$ – density of electron gas and $a_0$ determines the conventional Bohr radius. For the high density $(r_s < 1)$ limit the main state energy has the following form [2]:

$$\frac{1}{N} E_0 = \left( \frac{2.21}{r_s^2} - \frac{0.916}{r_s} + 0.0622 \ln r_s - 0.094 \right) \text{Ry}. \tag{1}$$

The first term corresponds to the kinetic energy, the second one determines the contribution of the exchange energy. The last two terms in (1) describe the correlation energy. It is necessary to note that in [3] another numerical coefficient before the logarithmic term was obtained

$$\frac{1}{N} E_{corr} = \left( 0.0570 \ln r_s - 0.094 \right) \text{Ry}. \tag{2}$$

Besides this observation we want to mark papers [4-6], where the different numerical values for the constant term figuring in (2) are presented. So, the question related to the correct evaluation of expression for the correlation energy remains open. The second task is related to the problem of the accurate (error controllable) calculations of the corrections to the correlation energy having more higher degrees of $r_s$. In paper [1] the authors suggested only a way in finding of the desired corrections proportional to $r_s$ and $r_s \ln(r_s)$, correspondingly. But the necessary calculations were not realized. In paper [7] in the frame of the random-phase approximation of Rayleigh-Schrödinger perturbation theory similar calculations were performed but it was shown that the numerical coefficient before $r_s$ can be evaluated only at $r_s \approx 1$. The desired terms having higher orders of $r_s$ were not found. Resuming the saying above one can conclude that the conventional calculations (in the frame of the diagram summation or Green function method with uncontrollable decoupling) of any term of the higher order entering into the correlation energy represents itself the complex and laborious work. So, it is necessary to develop a method which allows to realize these calculations for a wide class of interactions and temperatures having only one (and error controllable) approximation.

One of the authors of this paper (RRN) suggested the desired method, which was determined as the static fluctuations approximation (SFA). The SFA generalizes the well-known mean field method and allows to calculate the desired equilibrium correlation functions with necessary thermodynamic values of the many-body system considered. The basic idea of the SFA can be formulated as follows. Instead of replacing the components of the local field operator by its corresponding mean value it is possible to conserve its operator structure and find the spectrum of the remaining local field operator self-consistently. We need to clarify the basic idea of the SFA with the help of simple example. Let us consider some Hamiltonian describing interacting Fermi-gas. The simplest equation of motion for the Fermi operator $a_k^+(\tau)$, $\tau = it$ takes the form

$$\frac{da_k^+}{d\tau} = \left[H, a_k^+\right] = E_k a_k^+, \tag{3}$$

where $k$ (**k**,s) determines the state of a Fermi particle. The local field operator as a projection on states of a single Fermi particle is calculated easily from the relationship

$$E_k = \left[\left[H, a_k^+\right], a_k\right]_+, \quad [H, E_k] \cong 0. \tag{4}$$

We suppose also that this local field operator commutes with the Hamiltonian of the system considered. Simple calculations lead to the equation of the type

$$\langle n_k A\rangle = \left\langle \frac{A}{1+\exp(\beta E_k)}\right\rangle, \quad \beta = 1/T. \tag{5}$$

Here $E_k$ is the local field operator, which is defined by relationship (4). Operator $A$ in (5) determines an arbitrary combination of Fermi operators which commutes at least with operator $n_k = a_k^+ a_k$ belonging to the individual state $k$. The next step is to replace the eigen-values of the spectrum of the operator $E_k$ by their approximate values, which, in turn, are calculated self-consistently. Let us suppose that the spectrum of the local field operator $E_k$ is degenerate and finite. In this case it can be presented in the form

$$\prod_{s=1}^{n}(\lambda_s - E_k) = 0. \tag{6}$$

For this case using the Ceyley-Hamilton (CH) theorem [8] one can express any operator function $F(E_k)$ in the form of a finite polynomial

$$F(E_k) = \sum_{p=0}^{n-1}\eta_p (E_k)^p. \tag{7}$$

The unknown coefficients $\eta_p$ figuring in (7) are found from the system of linear equations

$$F(\lambda_s) = \sum_{p=0}^{n-1} \eta_p (\lambda_s)^p, \; s = 1, 2, ..., n. \qquad (8)$$

We are limited ourselves by consideration only linear term with respect to the difference operator $\Delta E_k = E_k - \langle E_k \rangle$. We suppose also that this operator satisfies approximately to the relationship

$$(\Delta E_k)^2 = a_k (\Delta E_k) + b_k,$$
$$a_k = \frac{\overline{(\Delta E_k)^3}}{\overline{(\Delta E_k)^2}}, \; b_k = \overline{(\Delta E_k)^2}. \qquad (9)$$

So, these expressions contain the basic approximation of the SFA. The unknown parameters $a_k$ and $b_k$ from (9) are calculated self-consistently, because they can be expressed by means of desired combination of operators $n_{k'}, k' \neq k$ figuring in values $a_k$ and $b_k$ and entering into operator $A$ in Eqn. (5). The details of calculations (when $a_k = 0$) are given below.

Using the CH theorem we find approximate relationship connecting operator $n_k$ with $\Delta E_k$. This equation has the form

$$\langle n_k A \rangle = \eta_0(T) \langle A \rangle + \eta_1(T) \langle \Delta E_k A \rangle. \qquad (10)$$

and is defined as far-distance equation (FDA) [9]. This equation allows to close all relationships for desired equilibrium correlation functions and thermodynamic values that need to be calculated. We do not give here the values of the temperature constants $\eta_0(T)$ and $\eta_1(T)$. They are easily obtained from the system (7) and equation (5). Other necessary details are given in the next section.

The SFA was applied with success to consideration of equilibrium properties of the Ising model of an arbitrary dimension [10,11,12], thermodynamics of the interacting of Bose-gas [13] and Habbard model [14]. For the justice sake it is necessary to note the fact that after close contacts of one of the authors (RRN) in Jordan with Prof. H. Ghassib he and his students started to use this new method for consideration of the thermodynamic properties of different Bose-systems [15-17].

The SFA essentially improves the mean-field approximation and allows to consider the thermodynamics of the strongly interacting systems and a wide interval of temperatures and fields. From the mathematical point of view we obtain the nonlinear system of difference equations (lattice models) or the closed system of nonlinear integral equations (Fermi and Bose systems) and so the further information can be obtained only by means of numerical methods or in the form of approximate decompositions with the usage of some small decomposition parameter. In this paper using the SFA it becomes possible to calculate the decomposition coefficients for correlation energy using $r_s$ (this parameter was defined above) as a small parameter for detailed consideration of the high density electron gas.

## 2. THE BASIC EQUATIONS

The complete Hamiltonian of the interacting homogeneous electron gas in the second quantization representation can be written as:

$$H = \sum_{\mathbf{k}s} \varepsilon_{\mathbf{k}} n_{\mathbf{k}s} + \frac{1}{2V} \sum_{\mathbf{q} \neq 0} V(\mathbf{q})(\rho_{\mathbf{q}}\rho_{-\mathbf{q}} - N), \tag{11}$$

$$\rho_{\mathbf{q}} = \frac{1}{\sqrt{N}} \sum_{\mathbf{k}s} a^{+}_{\mathbf{q}+\mathbf{k}s} a_{\mathbf{k}s}, \tag{12}$$

$$V(\mathbf{q}) = \frac{4\pi}{q^2 + q_0^2}, \tag{13}$$

here $\varepsilon_{\mathbf{k}} = k^2/2 - \mu$, $q_0 = \sqrt{\frac{6\pi n}{\varepsilon_F^0}} = \sqrt[3]{\frac{9}{\pi^2}} \frac{1}{a_0 r_s}$ - the screening Thomas-Fermi inverse radius. For convenience of the further calculations we realize the transformation of the given coordinate system in accordance with expression $\zeta = (3/4\pi)^{1/3} \mathbf{r}/r_0$. Then the Fermi-system of the given volume $V$ is transformed to the system with the volume $N$. The mean particle density in this space is equaled to unity. The Hamiltonian of the electron gas in the second quantization representation is transformed to the following form:

$$H' = \sum_{\mathbf{k}s} \varepsilon_{\mathbf{k}} n_{\mathbf{k}s} + \frac{\lambda}{2} \sum_{\mathbf{q} \neq 0} V(\mathbf{q})(\rho_{\mathbf{q}}\rho_{-\mathbf{q}} - 1), \tag{14}$$

where $\lambda = (4\pi/3)^{1/3} r_s$ defines the bond constant characterizing the intensity of interaction; the energy at this type of transformation is expressed in double rydbergs, The transition to the initial Hamiltonian $H$ is realized in accordance with relationship

$$H = \frac{1}{\lambda^2} H'. \tag{15}$$

We should note also that for the modified system the inverse screening radius is equaled to $q_0 = \sqrt{4k_F \lambda/\pi}$, and Fermi vector is becoming dimensionless and equaled $k_F = (3\pi^2)^{1/3}$. The equation of motion for the operator $a_{\mathbf{k}s}^+$ in the Heisenberg representation accepts the form:

$$\frac{da_{\mathbf{k}s}^+}{d\tau} = [H', a_{\mathbf{k}s}^+] = \varepsilon_k a_{\mathbf{k}s}^+ + \frac{\lambda}{2\sqrt{N}} \sum_{\mathbf{q}} V(\mathbf{q}) \left( \rho_{\mathbf{q}} a_{\mathbf{k}-\mathbf{q}s}^+ + a_{\mathbf{k}+\mathbf{q}s}^+ \rho_{-\mathbf{q}} \right). \qquad (16)$$

As before, $\tau = it$. We suppose that the Hamiltonian of the system can be presented in the form:

$$H' = \sum_{\mathbf{k}s} E_{\mathbf{k}s} n_{\mathbf{k}s}, \qquad (17)$$

where $E_{\mathbf{k}s}$ determines the local field operator and it can be calculated from Eqn. (4)

$$\frac{da_{\mathbf{k}s}^+}{d\tau} = E_{\mathbf{k}s} a_{\mathbf{k}s}^+ = \varepsilon_k a_{\mathbf{k}s}^+ + \frac{\lambda}{2\sqrt{N}} \sum_{\mathbf{q}} V(\mathbf{q}) \left( \rho_{\mathbf{q}} a_{\mathbf{k}-\mathbf{q}s}^+ + a_{\mathbf{k}+\mathbf{q}s}^+ \rho_{-\mathbf{q}} \right) \Rightarrow \qquad (18)$$

$$E_{\mathbf{k}s} = \left[ \frac{da_{\mathbf{k}s}^+}{d\tau}, a_{\mathbf{k}s} \right]_+ = \varepsilon_k - \frac{\lambda}{N} \sum_{\mathbf{q} \neq \mathbf{k}} V(\mathbf{k}-\mathbf{q}) n_{\mathbf{q}s}. \qquad (19)$$

Expression (19) determines the so-called Hartree-Fock operator, which satisfies to the following commutation rules

$$[E_{\mathbf{k}s}, a_{\mathbf{k}s}] = 0, \quad [H', E_{\mathbf{k}s}] = 0. \qquad (20)$$

In the mean field approximation the operator $E_{\mathbf{k}s}$ is replaced by its mean value and it means that the fluctuations of the local filed operator are not taken into account (Hartree-Fock approximation). Here we want to show *how* to take into account the fluctuations of the local filed operator. For this purpose we replace the square of the local filed displacement operator $\Delta E_{\mathbf{k}s} = E_{\mathbf{k}s} - <E_{\mathbf{k}s}>$ by its mean value. So, in accordance with expressions (9) written above we consider only the simplest case

$$\Delta E_{\mathbf{k}s}^2 \approx <\Delta E_{\mathbf{k}s}^2> \equiv b_{\mathbf{k}s} = \varphi_{\mathbf{k}s}^2. \qquad (21)$$

The approximate equality (21) is the key point of the SFA. The physical meaning of the SFA is that alongside with the mean value of the local field operator its quadratic fluctuations are taken into account. Then we want to show how to apply this basic approximation to calculation of the desired characteristics of the interacting Fermi-gas. For this purposes we find the expression for operator $a_{\mathbf{k}s}^+(\tau)$ in the Heisenberg representation

$$a^+_{\mathbf{k}s}(\tau) = a^+_{\mathbf{k}s} e^{E_{\mathbf{k}s}\tau}. \tag{22}$$

Then we determine the following correlation function

$$<a^+_{\mathbf{k}s}(\tau) a_{\mathbf{k}s} A> = <a^+_{\mathbf{k}s} a_{\mathbf{k}s} e^{E_{\mathbf{k}s}\tau}>. \tag{23}$$

Putting here $\tau = \beta$ ($\beta$ as before determines the inverse temperature) and taking into account the quantum identity for pair of noncommuting operators

$$<O_1(\beta)O_2> = <O_2 O_1>, \tag{24}$$

where $O_1(\tau)$ - determines an arbitrary operator in the Heisenberg representation and $O_2$ – is another arbitrary operator at $\tau = 0$, we obtain the relationship

$$<a^+_{\mathbf{k}s}(\beta) a_{\mathbf{k}s} A> = <a_{\mathbf{k}s} a^+_{\mathbf{k}s} A> = <a^+_{\mathbf{k}s} a_{\mathbf{k}s} e^{\beta E_{\mathbf{k}s}} A>. \tag{25}$$

Here $A$ - is an arbitrary combination of Fermi-operators commuting with operators $n_{\mathbf{k}s}$ and $E_{\mathbf{k}s}$. Taking into account the commutation rules for Fermi-operators one can present Eqn. (25) in the form

$$<n_{\mathbf{k}s}(1 + e^{\beta E_{\mathbf{k}s}}) A> = <A>. \tag{26}$$

Making a replacement $A \to (1 + e^{\beta E_{\mathbf{k}s}})^{-1} A$ one can rewrite Eqn.(26) in the following form

$$<n_{\mathbf{k}s} A> = <\frac{1}{1 + e^{\beta E_{\mathbf{k}s}}} A> = <f(E_{\mathbf{k}s}) A>, \tag{27}$$

where $f(x)$ determines the conventional Fermi-Dirac function. Equation (27) separating the Fermi-operators $n_{\mathbf{k}s}$ describing the state of a single particle from the components of the local field operator we determine as the far-distance equation (FDE) [9]. As it will be shown below the FDE allows to establish the desired relationships between equilibrium correlation functions of any order and thereby to close the system of nonlinear and self-consistent equations for the given many-body system considered. Using the CH theorem and taking into account relationships (7) and (8) we obtain

$$f(E_{\mathbf{k}s}) = \frac{1}{1 + e^{\beta E_{\mathbf{k}s}}} = \frac{1}{1 + e^{\beta(<E_{\mathbf{k}s}> + \Delta E_{\mathbf{k}s})}} = \eta_0(\mathbf{k}s) + \eta_1(\mathbf{k}s) \Delta E_{\mathbf{k}s}, \tag{28}$$

where

$$\eta_0(\mathbf{k}s) = \frac{1}{2}\big(f(<E_{\mathbf{k}s}> + \varphi_{\mathbf{k}s}) + f(<E_{\mathbf{k}s}> - \varphi_{\mathbf{k}s})\big), \tag{29}$$

$$\eta_1(\mathbf{k}s) = \frac{1}{2\varphi_{\mathbf{k}s}}\left(f(<E_{\mathbf{k}s}>+\varphi_{\mathbf{k}s}) - f(<E_{\mathbf{k}s}>-\varphi_{\mathbf{k}s})\right). \tag{30}$$

In the result of this decomposition the FDE (28) becomes more convenient for further analysis

$$<n_{\mathbf{k}s}A> = \eta_0(\mathbf{k}s)<A> + \eta_1(\mathbf{k}s)<\Delta E_{\mathbf{k}s}A>, \tag{31}$$

Here we introduce the operator

$$\Delta E_{\mathbf{k}s} = -\frac{\lambda}{N}\sum_{\mathbf{q}\neq\mathbf{k}}V(\mathbf{k}-\mathbf{q})\Delta n_{\mathbf{q}s}. \tag{32}$$

Equation (31) allows to obtain the closed nonlinear system of integral equations for calculation of the desired thermodynamic values of the Fermi-system considered. Putting in (31) $A=1$ we obtain the equation for calculation of $<n_{\mathbf{k}s}>$:

$$<n_{\mathbf{k}s}> = \eta_0(\mathbf{k}s). \tag{33}$$

Using (33) we can rewrite FDE (31) in an elegant form that it is more convenient for further calculations

$$<\Delta n_{\mathbf{k}s}A> = \eta_1(\mathbf{k}s)<\Delta E_{\mathbf{k}s}A>. \tag{34}$$

For calculation of pair correlation functions we put in (34) $A=\Delta n_{\mathbf{q}s'}$ and take into account the fact that in (34) the wave vectors $\mathbf{q}\neq\mathbf{k}$ cannot coincide with each other. In the result of this substitution we obtain another equation

$$<\Delta n_{\mathbf{k}s}\Delta n_{\mathbf{q}s'}> = \Delta_{\mathbf{k}s}\delta_{\mathbf{k},\mathbf{q}}\delta_{s,s'} - \frac{\lambda\eta_1(\mathbf{k}s)}{N}\sum_{\mathbf{q}'}V(\mathbf{k}-\mathbf{q}')<\Delta n_{\mathbf{q}'s}\Delta n_{\mathbf{q}s'}>\left(1-\delta_{\mathbf{k},\mathbf{q}}\delta_{s,s'}\right), \tag{35}$$

where we introduced the notations $\Delta_{\mathbf{k}s} = <n_{\mathbf{k}s}>(1-<n_{\mathbf{k}s}>)$ and take into account the expression (32) for $\Delta E_{\mathbf{k}s}$. Besides this we used the kinematic identity that is correct for any Fermi-system $<(\Delta n_{\mathbf{k}s})^2> = \Delta_{\mathbf{k}s} = <n_{\mathbf{k}s}>(1-<n_{\mathbf{k}s}>)$. Equation (35) written in the form of the integral equation serves for calculation of the binary correlation functions for the Fermi-system considered. Then it is necessary to obtain the equation for calculation of the local field quadratic fluctuations. For this purpose we put in the FDE (34) $A=\Delta E_{\mathbf{k}s}$ and using the definition (21) we obtain

$$\varphi_{\mathbf{k}s}^2 = -\frac{\lambda}{\eta_1(\mathbf{k}s)N} \sum_{\mathbf{q} \neq \mathbf{k}} V(\mathbf{k} - \mathbf{q}) < \Delta n_{\mathbf{k}s} \Delta n_{\mathbf{q}s} >. \tag{36}$$

The set of equations (19), (33), (35) and (36) forms the enclosed system of nonlinear integral equations for finding of the desired values $< E_{\mathbf{k}s} >$, $< n_{\mathbf{k}s} >$, $< \Delta n_{\mathbf{k}s} \Delta n_{\mathbf{q}s'} >$, $\varphi_{\mathbf{k}s}$.

In the frame of the SFA it is easy to obtain the expressions for mean energy and partition function (here the mean energy is measured in rydbergs).

$$< H > = \frac{2}{\lambda^2} < H' > = \frac{2}{\lambda^2} \sum_{\mathbf{k}s} < n_{\mathbf{k}s} E_{\mathbf{k}s} > = \frac{2}{\lambda^2} \sum_{\mathbf{k}s} \left( \eta_0(\mathbf{k}s) < E_{\mathbf{k}s} > + \eta_1(\mathbf{k}s) \varphi_{\mathbf{k}s}^2 \right), \tag{37}$$

$$\ln Q = \ln \mathrm{Sp}\, e^{-\beta H} = \ln \mathrm{Sp} \exp\left(-\beta \frac{2}{\lambda^2} \sum_{\mathbf{k}s} E_{\mathbf{k}s} n_{\mathbf{k}s}\right) = \ln \prod_{\mathbf{k}s} \sum_{n_{\mathbf{k}s}, \Delta E_{\mathbf{k}s}} \exp(-\beta \frac{2}{\lambda^2} E_{\mathbf{k}s} n_{\mathbf{k}s}) =$$

$$= N \ln 2 + \sum_{\mathbf{k}s} \ln(1 + \exp(-\beta \frac{2}{\lambda^2} < E_{\mathbf{k}s} >) \mathrm{ch}\left(\beta \frac{2}{\lambda^2} \varphi_{\mathbf{k}s}\right). \tag{38}$$

We want to stress here again that the closed systems of equations are based only on the single approximation (21) and in the limit of this approximation one can describe the Fermi-system in wide range of temperatures and potentials. A possible generalization of these equations that takes into account the influence of the parameter of asymmetry $a_k$ can be considered in other papers.

## 3. SOLUTION OF INTEGRAL EQUATIONS FOR THE HIGH DENSITY ELECTRON GAS

For a possibility of obtaining of analytical solutions we consider interacting electron gas with high density. It is easy to notice that the high density electron gas ($r_s \to 0$) corresponds to weak interaction in (14) in comparison with kinetic energy. It means that the fluctuations of the local field are small in comparison with mean value of this field $\varphi_{\mathbf{k}s} << < E_{\mathbf{k}s} >$. So, in this case for the coefficient $\eta_1(\mathbf{k}s)$ from (30) one can derive the following approximate expression

$$\eta_1(\mathbf{k}s) \approx f'(< E_{\mathbf{k}s} >_0) = -\beta f(< E_{\mathbf{k}s} >_0) f(- < E_{\mathbf{k}s} >_0), \tag{39}$$

where $<E_{\mathbf{k}s}>_0 = <E_{\mathbf{k}s}>|_{\varphi_{\mathbf{k}s}=0}$ does not depend on the value of $\varphi_{\mathbf{k}s}$ and, hence, from the influence of the binary correlation function $<\Delta n_{\mathbf{k}s}\Delta n_{\mathbf{q}s'}>$. For this case one can apply for solution of Eqn. (35) the step-by-step method and then inserting this result into (36) we get finally the desired decomposition

$$\varphi_{\mathbf{k}s}^2 = \frac{1}{N}\sum_{\mathbf{q}\neq\mathbf{k}}\Delta_{\mathbf{q}s}V(\mathbf{k}-\mathbf{q})\sum_{n=1}^{\infty}\left(\frac{\lambda}{N}\right)^{n+1}\beta^{n-1}\left(\prod_{i=1}^{n-1}f(<E_{\mathbf{k}s}>_0)f(-<E_{\mathbf{k}s}>_0)\right)\times$$
$$\times V(\mathbf{k}-\mathbf{k}_1)V(\mathbf{k}_1-\mathbf{k}_2)...V(\mathbf{k}_{n-1}-\mathbf{q}). \qquad (40)$$

Taking into account the small value of $\varphi_{\mathbf{k}s}$ one can decompose the equation (33) and present it in the from of the infinite series containing the integer degrees of $\varphi_{\mathbf{k}s}$:

$$<n_{\mathbf{k}s}> = f(<E_{\mathbf{k}s}>_0) + \frac{\beta^2}{2}f_3(<E_{\mathbf{k}s}>_0)\varphi_{\mathbf{k}s}^2 + ..., \qquad (41)$$

where $f_n(x) = (-1)^{n-1}f^{(n)}(x)$ and $f(x)$ determines the conventional Fermi-Dirac function. Equations (40) and (41) constitute the closed system of equations for the finding of the mean value $<n_{\mathbf{k}s}>$ and $\varphi_{\mathbf{k}s}$. This system can be solved also by the step-by-step method with given accuracy over the parameter $\lambda$. Keeping only the terms of the order $\varphi_{\mathbf{k}s}^2$ we get approximately

$$<H> = \frac{1}{\lambda^2}\sum_{\mathbf{k}s}\left(<n_{\mathbf{k}s}>_0<E_{\mathbf{k}s}>_0 + \left[\frac{\beta^2}{2}<E_{\mathbf{k}s}>_0 f_3(<E_{\mathbf{k}s}>_0) - \beta f_2(<E_{\mathbf{k}s}>_0)\right]\varphi_{\mathbf{k}s}^2 - \right.$$
$$\left. -\lambda\beta^2 <n_{\mathbf{k}s}>_0 \frac{1}{2N}\sum_{\mathbf{q}\neq\mathbf{k}}V(\mathbf{k}-\mathbf{q})f_3(<E_{\mathbf{q}s}>_0)\varphi_{\mathbf{q}s}^2\right), \qquad (42)$$

where $<n_{\mathbf{k}s}>_0 = f(<E_{\mathbf{k}s}>_0)$.

## 4. THE ENERGY OF THE GROUND STATE AND HEAT CAPACITY OF THE HIGH-DENSITY ELECTRON GAS AT ZERO TEMPERATURE

In the general expressions obtained for the high density electron gas ($r_s \to 0$), we keep the terms up to $\lambda^2$. Using the decomposition (40) we present it in the following approximate form

$$\varphi_{\mathbf{k}s}^2 \approx \left(\frac{\lambda}{N}\right)^2 \sum_{\mathbf{q}\neq\mathbf{k}} V^2(\mathbf{k}-\mathbf{q}) f_2(\varepsilon_{\mathbf{q}}). \tag{43}$$

It is necessary to note also that from expression

$$\frac{1}{N}\sum_{\mathbf{k}}\ldots = \frac{1}{(2\pi)^3}\int\ldots d^3k, \tag{44}$$

it follows that values for quadratic fluctuations $\varphi_{\mathbf{k}s}^2$ proportional to $1/N$ and becomes negligible, especially for 3D-case. This is expected result because we are solving the nonlinear system of integral equations at high density limit when $\varphi_{\mathbf{k}s} \ll \langle E_{\mathbf{k}s}\rangle$. But in principle the closed system of equations obtained in the second part of this work is correct for any relationship between $\varphi_{\mathbf{k}s}$ and $\langle E_{\mathbf{k}s}\rangle$. For example, for the low density case ($r_s \to \infty$), when potential energy exceeds the kinetic energy the values of fluctuations $\varphi_{\mathbf{k}s}$ become comparable with mean value of the local field $\langle E_{\mathbf{k}s}\rangle$

It is necessary to note also that in expression (17) in projection of the given Hamiltonian $H'$ to the number of states $n_{\mathbf{k}s}$ some terms in the expression for mean energy are not taken into account. These terms arise because of the fact that some terms do not commute with Fermi operators $a_{\mathbf{k}s}^+(a_{\mathbf{k}s})$. For the given case this term coincides with exchange (or background) energy. This term is determined by expression

$$\frac{1}{N}E_{ex} = -\frac{2}{\lambda N}\sum_{\mathbf{q}} V(\mathbf{q}-\mathbf{q}_0). \tag{45}$$

Following the results of the book [18] one can show that this term proportional to $-0.916/r_s$.

Taking into account the above saying remarks and decomposing the mean value $<n_{\mathbf{k}s}>_0$ in (41) up to $\lambda^2$ term one can get the desired expression for the ground state energy. The calculation scheme and evaluation of some terms are presented in the Mathematical Appendix I. The final expression for the energy of the ground state (at $T=0$) can be presented in the form

$$\frac{1}{N}E_0 = \left(\frac{2.21}{r_s^2} - \frac{0.916}{r_s} + \frac{0.781}{\sqrt{r_s}} + 1.556\sqrt{r_s} + 0.304\ln r_s - 1.002 + O(r_s)\right) \text{Ry}. \tag{46}$$

Here we did not take into account the temperature dependence of the chemical potential $\mu(T) \approx \mu(0)$. Comparing (1) with (46) one can notice that formally they are strongly deviated from each other. But being plotted for comparison on the same figure (see Fig.1) we see that they practically coincide with each other and the relative fitting error does not exceed 0.5%. This fact says in favor of very good coincidence of these expressions if we take into account the fact that in papers [3-6] the different values for the constant and the coefficient before logarithmic term are given. In Fig.2. we show the relative differences between expressions obtained by other authors [1,3,6] and expression obtained in the frame of SFA. We compare the previous expressions with expression (46) having in mind only one reason. The expressions obtained by other authors contain uncontrollable errors (because of separate summation of a certain class of diagrams). In our case we made only one supposition (21) and this supposition is controllable and the ratio $\varphi_{\mathbf{k}s} / \langle E_{\mathbf{k}s} \rangle \ll 1$.

Taking the derivative from the mean energy with respect to temperature one can get expression for heat capacity at low temperatures ($T = 0$). For comparison it is convenient to present this result in the form

$$\frac{C_F}{C} = 1 + 0.055 r_s - 0.070 r_s^{3/2} + O(r_s^2). \tag{47}$$

where $C_F$ determines the heat capacity of the non-interacting Fermi-gas. For comparison we are giving our result in comparison of similar results obtained by other authors [19, 20]. They are presented on Fig.3. Here we observe more essential discrepancies in comparison with results presented in Fig. 1. But again we consider our result as more accurate because it contains the minimal value of the error expressed by Eqn. (21). In the frame of the SFA method it is rather easy to find the terms of higher orders with respect to $r_s$. For this aim it is sufficient in composition for mean energy in expression containing $<n_{\mathbf{k}s}>_0$ to keep not only quadratic terms proportional to $\lambda^2$. If in this $\lambda$- decomposition one keeps the terms proportional to $\lambda^3$ then one can find the terms of the order $r_s$ и $r_s^2$ in expressions for mean energy and heat capacity, correspondingly. Additional terms which are appeared in decomposition for mean energy are given in the Mathematical Appendix I. For this case expressions for the ground state energy and heat capacity accept the forms

$$\frac{1}{N} E_0 = \left( \frac{2.21}{r_s^2} - \frac{0.916}{r_s} + \frac{0.781}{\sqrt{r_s}} + 1.556\sqrt{r_s} + 0.304 \ln r_s - 1.002 + \right.$$

$$+(-1.644+0.538\ln(r_s))r_s+O(r_s^{3/2})\Big)\text{Ry}, \tag{48}$$

$$\frac{C_F}{C}=1+0.055r_s-0.070r_s^{3/2}+$$

$$+(-0.026+0.012\ln(r_s)-0.007\ln^2(r_s))r_s^2+O(r_s^2). \tag{49}$$

The contributions of the third order terms with respect to expression (46) are shown on Fig.4. One can notice that the third order terms are becoming more essential with increasing of the value of parameter $r_s$. The same tendency is observed in comparison of expressions (47) with (49) for heat capacity. The corresponding plots are presented by Fig. 5.

## 5. RESULTS AND DISCUSSION

In this paper we demonstrated the calculation of the correlation energy in the frame of the SFA method. As one can notice from the content of this paper the calculation scheme is very simple but leads finally to the solution of the nonlinear system of the integral equations. This system of equations based on only one supposition (21) allows easily to calculate the desired correlation functions and obtain the expressions for the ground state energy and heat capacity that are in accordance with results obtained by other authors. It is interesting to note that the simplest decoupling scheme (21) admits some generalization in relationship (9). More general scheme (9) allows considering the asymmetrical spectrum of the local field operator and opens new possibilities in receiving more general results. This generalization for Fermi-gas and other many-body systems with strong interaction between particles merits the separate research.

# MATHEMATICAL APPENDIX I

In order to find an analytical expression for mean energy it is necessary in expression (42) to realize the decomposition of expression $<n_{\mathbf{k}s}>_0$ with respect to parameter $\lambda$ and take into account the fact that fluctuation terms $\varphi_{\mathbf{k}s}$ in the case of high density proportional to $1/N$. In this case keeping in the corresponding decomposition the terms proportional to $\lambda^3$ expression for $<n_{\mathbf{k}s}>_0$ takes the form

$$<n_{\mathbf{k}s}>_0 = f(\varepsilon_{\mathbf{k}}) + \frac{\lambda}{N}\beta f_2(\varepsilon_{\mathbf{k}}) \sum_{\mathbf{q}\neq\mathbf{k}} V(\mathbf{k}-\mathbf{q})f(\varepsilon_{\mathbf{q}}) +$$

$$+ \left(\frac{\lambda}{N}\right)^2 \beta^2 f_2(\varepsilon_{\mathbf{k}}) \left[ \left(\frac{1}{2}-f(\varepsilon_{\mathbf{k}})\right)\left(\sum_{\mathbf{q}\neq\mathbf{k}} V(\mathbf{k}-\mathbf{q})f(\varepsilon_{\mathbf{q}})\right)^2 + \sum_{\mathbf{q}\neq\mathbf{k}} V(\mathbf{k}-\mathbf{q})f_2(\varepsilon_{\mathbf{q}}) \sum_{\mathbf{q}_1\neq\mathbf{q}} V(\mathbf{q}-\mathbf{q}_1)f(\varepsilon_{\mathbf{q}_1}) \right]$$

$$+ \left(\frac{\lambda}{N}\right)^3 \beta^3 \left[ \frac{1}{6} f_4(\varepsilon_{\mathbf{k}}) \left(\sum_{\mathbf{q}\neq\mathbf{k}} V(\mathbf{k}-\mathbf{q})f(\varepsilon_{\mathbf{q}})\right)^3 + \right.$$

$$+ f_3(\varepsilon_{\mathbf{k}}) \sum_{\mathbf{q}\neq\mathbf{k}} V(\mathbf{k}-\mathbf{q})f(\varepsilon_{\mathbf{q}}) \sum_{\mathbf{q}\neq\mathbf{k}} V(\mathbf{k}-\mathbf{q})f_2(\varepsilon_{\mathbf{q}}) \sum_{\mathbf{q}\neq\mathbf{q}_1} V(\mathbf{q}-\mathbf{q}_1)f(\varepsilon_{\mathbf{q}_1}) +$$

$$+ \frac{1}{2} f_2(\varepsilon_{\mathbf{k}}) \sum_{\mathbf{q}\neq\mathbf{k}} V(\mathbf{k}-\mathbf{q})f_3(\varepsilon_{\mathbf{q}}) \left(\sum_{\mathbf{q}\neq\mathbf{q}_1} V(\mathbf{q}-\mathbf{q}_1)f(\varepsilon_{\mathbf{q}_1})\right)^2 +$$

$$\left. + f_2(\varepsilon_{\mathbf{k}}) \sum_{\mathbf{q}\neq\mathbf{k}} V(\mathbf{k}-\mathbf{q})f_2(\varepsilon_{\mathbf{q}}) \sum_{\mathbf{q}\neq\mathbf{q}_1} V(\mathbf{q}-\mathbf{q}_1)f_2(\varepsilon_{\mathbf{q}_1}) \sum_{\mathbf{q}_2\neq\mathbf{q}_1} V(\mathbf{q}_1-\mathbf{q}_2)f(\varepsilon_{\mathbf{q}_2}) \right] \quad (AI.1)$$

Then the mean energy can be present in the form

$$\frac{1}{N}E = \frac{1}{N}<H> = \frac{2}{\lambda^2 N}<H'> = \frac{2}{\lambda^2}(E_0 + E_1 + E_2 + E_3), \qquad (AI.2)$$

here $E_0$ contains the zero order terms with respect to $\lambda$

$$E_0 = \frac{2}{N}\sum_{\mathbf{k}} \varepsilon_k f(\varepsilon_k), \qquad (AI.3)$$

The term $E_1$ contains only a pair of the first order terms with respect to $\lambda$

$$E_1 = E_1^{(1)} - E_1^{(2)}, \qquad (AI.4)$$

$$E_1^{(1)} = \frac{2\lambda\beta}{N^2}\sum_{\mathbf{k}}\varepsilon_k f_2(\varepsilon_k)\sum_{\mathbf{q}\neq\mathbf{k}}V(\mathbf{k}-\mathbf{q})f(\varepsilon_q), \tag{AI.5}$$

$$E_1^{(2)} = \frac{2\lambda}{N^2}\sum_{\mathbf{k}}f(\varepsilon_k)\sum_{\mathbf{q}\neq\mathbf{k}}V(\mathbf{k}-\mathbf{q})f(\varepsilon_q), \tag{AI.6}$$

Expression $E_2$ figuring in (AI.2) contains three terms of the second order with respect to $\lambda$

$$E_2 = E_2^{(1)} + E_2^{(2)} - E_2^{(3)}, \tag{AI.7}$$

$$E_2^{(1)} = \frac{\lambda^2\beta^2}{N^3}\sum_{k}\varepsilon_k f_3(\varepsilon_k)\left(\sum_{\mathbf{q}\neq\mathbf{k}}V(\mathbf{k}-\mathbf{q})f(\varepsilon_q)\right)^2, \tag{AI.8}$$

$$E_2^{(2)} = \frac{2\lambda^2\beta^2}{N^3}\sum_{k}\varepsilon_k f_2(\varepsilon_k)\sum_{\mathbf{q}\neq\mathbf{k}}V(\mathbf{k}-\mathbf{q})f_2(\varepsilon_q)\sum_{\mathbf{q}_1\neq\mathbf{q}}V(\mathbf{q}-\mathbf{q}_1)f(\varepsilon_{q_1}), \tag{AI.9}$$

$$E_2^{(3)} = \frac{4\lambda^2\beta}{N^3}\sum_{k}f(\varepsilon_k)\sum_{\mathbf{q}\neq\mathbf{k}}V(\mathbf{k}-\mathbf{q})f_2(\varepsilon_q)\sum_{\mathbf{q}_1\neq\mathbf{q}}V(\mathbf{q}-\mathbf{q}_1)f(\varepsilon_{q_1}). \tag{AI.10}$$

Expression $E_3$ from (AI.2) contains already 6 terms of the third order

$$E_3 = E_3^{(1)} + E_3^{(2)} + E_3^{(3)} + E_3^{(4)} - E_3^{(5)} - E_3^{(6)}, \tag{AI.11}$$

$$E_3^{(1)} = \frac{\lambda^3\beta^3}{3N^4}\sum_{\mathbf{k}}\varepsilon_\mathbf{k} f_4(\varepsilon_\mathbf{k})\left(\sum_{\mathbf{q}\neq\mathbf{k}}V(\mathbf{k}-\mathbf{q})f(\varepsilon_\mathbf{q})\right)^3, \tag{AI.12}$$

$$E_3^{(2)} = \frac{2\lambda^3\beta^3}{N^4}\sum_{\mathbf{k}}\varepsilon_\mathbf{k} f_3(\varepsilon_\mathbf{k})\sum_{\mathbf{q}'\neq\mathbf{k}}V(\mathbf{k}-\mathbf{q}')f(\varepsilon_{\mathbf{q}}{'})\times$$
$$\times\sum_{\mathbf{q}\neq\mathbf{k}}V(\mathbf{k}-\mathbf{q})f_2(\varepsilon_\mathbf{q})\sum_{\mathbf{q}\neq\mathbf{q}_1}V(\mathbf{q}-\mathbf{q}_1)f(\varepsilon_{\mathbf{q}_1}), \tag{AI.13}$$

$$E_3^{(3)} = \frac{\lambda^3\beta^3}{N^4}\sum_{\mathbf{k}}\varepsilon_\mathbf{k} f_2(\varepsilon_\mathbf{k})\sum_{\mathbf{q}\neq\mathbf{k}}V(\mathbf{k}-\mathbf{q})f_3(\varepsilon_\mathbf{q})\left(\sum_{\mathbf{q}\neq\mathbf{q}_1}V(\mathbf{q}-\mathbf{q}_1)f(\varepsilon_{\mathbf{q}_1})\right)^2, \tag{AI.14}$$

$$E_3^{(4)} = \frac{2\lambda^3\beta^3}{N^4}\sum_{\mathbf{k}}\varepsilon_\mathbf{k} f_2(\varepsilon_\mathbf{k})\sum_{\mathbf{q}\neq\mathbf{k}}V(\mathbf{k}-\mathbf{q})f_2(\varepsilon_\mathbf{q})\times$$
$$\times\sum_{\mathbf{q}_1\neq\mathbf{q}}V(\mathbf{q}-\mathbf{q}_1)f_2(\varepsilon_{\mathbf{q}_1})\sum_{\mathbf{q}_2\neq\mathbf{q}_1}V(\mathbf{q}_1-\mathbf{q}_2)f(\varepsilon_{\mathbf{q}_2}), \tag{AI.15}$$

$$E_3^{(5)} = \frac{\lambda^3 \beta^2}{N^4} \sum_{\mathbf{k}} f(\varepsilon_{\mathbf{k}}) \sum_{\mathbf{q} \neq \mathbf{k}} V(\mathbf{k}-\mathbf{q}) f_3(\varepsilon_{\mathbf{q}}) \left( \sum_{\mathbf{q}_1 \neq \mathbf{q}} V(\mathbf{q}-\mathbf{q}_1) f(\varepsilon_{\mathbf{q}_1}) \right)^2, \quad \text{(AI.16)}$$

$$E_3^{(6)} = \frac{2\lambda^3 \beta^2}{N^4} \sum_{\mathbf{k}} f(\varepsilon_{\mathbf{k}}) \sum_{\mathbf{q} \neq \mathbf{k}} V(\mathbf{k}-\mathbf{q}) f_2(\varepsilon_{\mathbf{q}}) \times$$

$$\times \sum_{\mathbf{q}_1 \neq \mathbf{q}} V(\mathbf{q}-\mathbf{q}_1) f_2(\varepsilon_{\mathbf{q}_1}) \sum_{\mathbf{q}_2 \neq \mathbf{q}_1} V(\mathbf{q}_1-\mathbf{q}_2) f(\varepsilon_{\mathbf{q}_2}). \quad \text{(AI.17)}$$

For understanding the basic calculation scheme we consider the calculation of the term $E_2^{(2)}$ (determined by expression (AI.9)) in detail. In the beginning we replace the summation by integration in accordance with expression (44) taking into account also definition (13) for potential $V(\mathbf{k}-\mathbf{q})$

$$E_2^{(2)} = \frac{2\lambda^2 \beta^2}{(2\pi)^9} \int \varepsilon_k f_2(\varepsilon_k) d^3k \int V(\mathbf{k}-\mathbf{q}) f_2(\varepsilon_q) d^3q \int V(\mathbf{q}-\mathbf{q}_1) f(\varepsilon_{q_1}) d^3 q_1 =$$

$$= \frac{\lambda^2 \beta^2}{\pi^4} \int_0^\infty \varepsilon_k f_2(\varepsilon_k) k^2 dk \int_{-1}^1 d\nu_1 \int_0^\infty \frac{f_2(\varepsilon_q) q^2 dq}{k^2 + q^2 - 2kq\nu_1 + q_0^2} \int_{-1}^1 d\nu_2 \int_0^\infty \frac{f(\varepsilon_{q_1}) q_1^2 dq_1}{q_1^2 + q^2 - 2q_1 q \nu_2 + q_0^2} =$$

$$= \frac{\lambda^2 \beta^2}{4\pi^4} \int_0^\infty \varepsilon_k f_2(\varepsilon_k) k dk \int_0^\infty f_2(\varepsilon_q) \ln \left| \frac{(k+q)^2 + q_0^2}{(k-q)^2 + q_0^2} \right| dq \times$$

$$\times \int_0^\infty f(\varepsilon_{q_1}) \ln \left| \frac{(q_1+q)^2 + q_0^2}{(q_1-q)^2 + q_0^2} \right| q_1 dq_1. \quad \text{(AI.18)}$$

Changing the variables $\frac{k^2}{2} = x$ $\frac{q^2}{2} = y$ $\frac{q_1^2}{2} = z$ we obtain

$$E_2^2 = \frac{\lambda^2 \beta^2}{4\sqrt{2}\pi^4} \int_0^\infty f_2\left(\frac{x-\mu}{T}\right) dx \int_0^\infty f_2\left(\frac{y-\mu}{T}\right) dy \int_0^\infty f_1\left(\frac{z-\mu}{T}\right) h(x,y,z) dz, \quad \text{(AI.19)}$$

where

$$h(x,y,z) = \frac{x}{\sqrt{y}} \ln\left( \frac{(\sqrt{x}+\sqrt{y})^2 + \varepsilon_0}{(\sqrt{x}-\sqrt{y})^2 + \varepsilon_0} \right) \ln\left( \frac{(\sqrt{y}+\sqrt{z})^2 + \varepsilon_0}{(\sqrt{y}-\sqrt{z})^2 + \varepsilon_0} \right), \quad \text{(AI.20)}$$

here $\varepsilon_0 = q_0^2/2$. Using relationships (AII.2) and (AII.3) that are given in the Mathematical Appendix II, we receive the following expression

$$E_2^2 = \frac{\lambda^2 \beta^2}{4\sqrt{2}\pi^4} \int_0^\infty f_2\left(\frac{x-\mu}{T}\right) dx \int_0^\infty f_2\left(\frac{y-\mu}{T}\right) dy \left\{ \int_0^\mu h(x,y,z)dz + \frac{\pi^2}{6} h_z'(x,y,\mu)T^2 \right\} =$$

$$= \frac{\lambda^2 \beta}{4\sqrt{2}\pi^4} \int_0^\infty f_2\left(\frac{x-\mu}{T}\right) dx \left\{ \int_0^\mu h(x,\mu,z)dz + \frac{\pi^2}{6}\left( h_z'(x,\mu,\mu) + \int_0^\mu h_{yy}''(x,\mu,z)dz \right) T^2 \right\} =$$

$$= \frac{\lambda^2}{4\sqrt{2}\pi^4} \left\{ \int_0^\mu h(\mu,\mu,z)dz + \right.$$

$$\left. + \frac{\pi^2}{6}\left( h_z'(\mu,\mu,\mu) + \int_0^\mu \left( h_{xx}''(\mu,\mu,z) + h_{yy}''(\mu,\mu,z) \right) dz \right) T^2 \right\} \quad (AI.21)$$

Realizing simple but cumbersome calculations we obtain finally

$$E_2^2 = \frac{\lambda^2}{4\sqrt{2}\pi^4} \left\{ \frac{p\varepsilon_0^{3/2}}{2}\left( p + \ln(1+p^2) - 2p\,\text{arctg}\,p \right)\ln(1+p^2) + \right.$$

$$\frac{\pi^2}{6\sqrt{\varepsilon_0}}\left( -\frac{2p(p^4-p^2-4)}{(1+p^2)^2} + \frac{4(p^2+2)(p^2-1)}{(1+p^2)^2} \text{arctg}\,p - \right.$$

$$\left.\left. -\frac{2(4p^4+9p^2+3)}{p(1+p^2)^2}\ln(1+p^2) + \frac{2(p^2+3)}{p^3}\ln^2(1+p^2) \right) T^2 + ... \right\}, \quad (AI.22)$$

where $p = 2\sqrt{\mu/\varepsilon_0}$. Other terms figuring in the basic expression (AI.2) are calculated by a similar way.

We want to remind here the following. All mean values for the desired thermodynamic functions were obtained with the use of the Hamiltonian $H'$. For transition to $H$ is it necessary to take into account the relationship

$$\langle \hat{A}(\beta) \rangle = Q^{-1} Sp\left( e^{-H\beta} \hat{A} \right) = Q^{-1} Sp\left( e^{-H'\frac{2}{\lambda^2}\beta} \hat{A} \right) = \left\langle \hat{A}\left(\frac{2}{\lambda^2}\beta\right) \right\rangle', \quad (AI.23)$$

and replacements

$$\beta \to \tilde{\beta} = \frac{2}{\lambda^2}\beta, \quad T \to \tilde{T} = \frac{\lambda^2}{2}T. \tag{AI.24}$$

**MATHEMATICAL APPENDIX II**

Let us consider a set of the following integrals

$$I_i(T,\mu) = \int_0^\infty g(x) f_i\left(\frac{x-\mu}{T}\right) dx, \quad i = 1, 2, 3, 4, \tag{AII.1}$$

where $g(x)$ defines a smooth function, $f_i(x) = (-1)^{i-1} f^{(i-1)}(x)$, and $f(x)$, in turn, determines the conventional Fermi-Dirac function and $f^{(i)}(x)$ its corresponding derivative. In the case $i = 1$ at low temperatures we have the well-known low temperature decomposition

$$I_1(T,\mu) = \int_0^\mu g(x) dx + \frac{\pi^2}{6} g'(\mu) T^2 + \dots. \tag{AII.2}$$

Differentiating the both parts of this expression with respect to chemical potential $\mu$ one can get useful expressions

$$I_2(T,\mu) = g(\mu) T + \frac{\pi^2}{6} g''(\mu) T^3 + \dots, \tag{AII.3}$$

$$I_3(T,\mu) = g'(\mu) T^2 + \frac{\pi^2}{6} g'''(\mu) T^4 + \dots, \tag{AII.4}$$

$$I_4(T,\mu) = g''(\mu) T^3 + \frac{\pi^2}{6} g^{(4)}(\mu) T^5 \dots \tag{AII.5}$$

that were used above in evaluation of expression (AI.21).

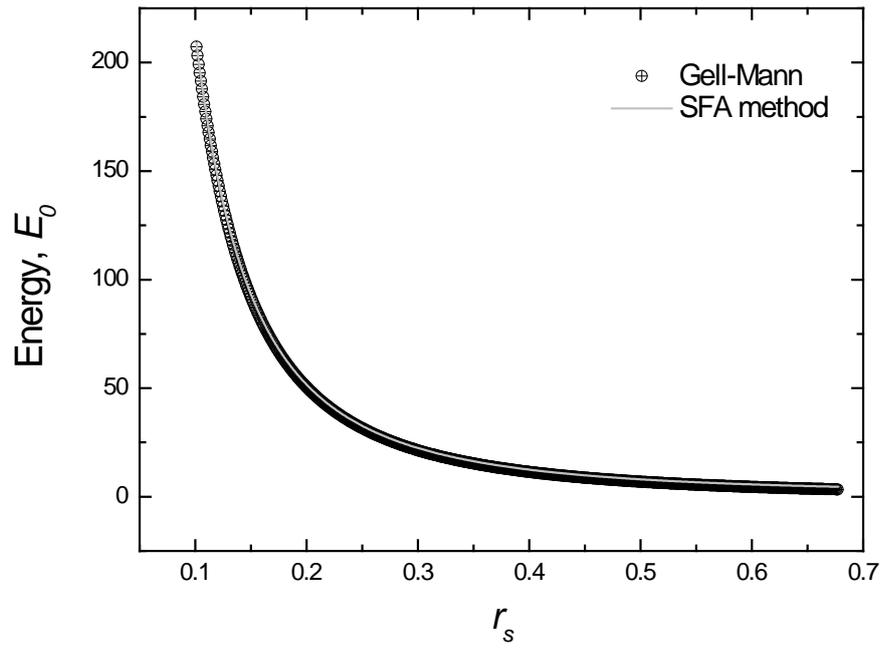

**Fig.1**. The comparison of expression for ground state energy obtained by Gell-Mann (1) with expression (46) ($E_0$) obtained in the frame of SFA. As one can notice from this comparison these expressions practically coincides with each other. The relative error (defined as the ratio of the standard deviation taken from the difference ($E_{SFA}$-$E_{GM}$) to $E_{SFA}$ mean value) does not exceed 0.5%.

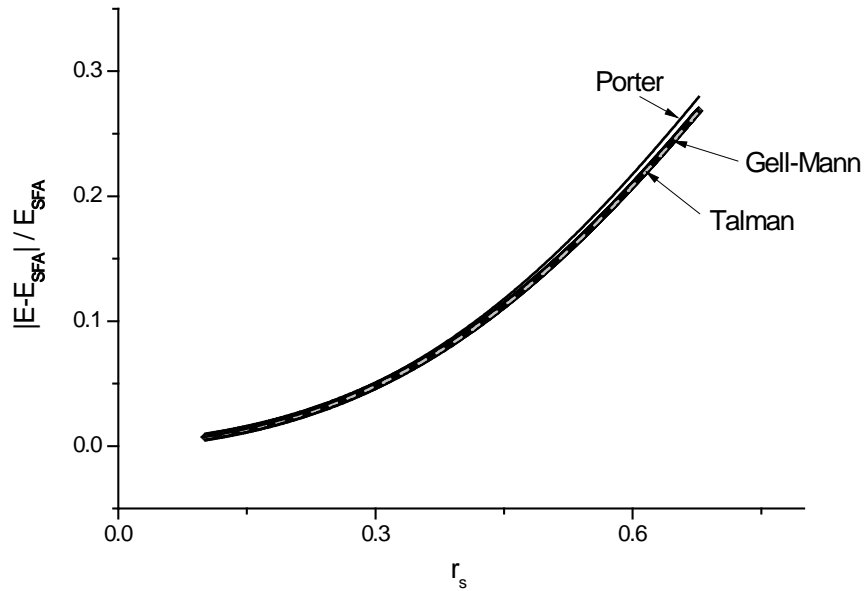

**Fig.2**. The relative expressions $|E_0 - E_{SFA}|/E_{SFA}$ for the ground energy obtained by Porter, Gell-Mann and Talman. The relative deviations obtained by two last authors practically coincide with each other. The visible deviations at relatively large $r_s$ are observed for the Porter's results. All relative differences are increasing with increasing of the ratio $r_s$.

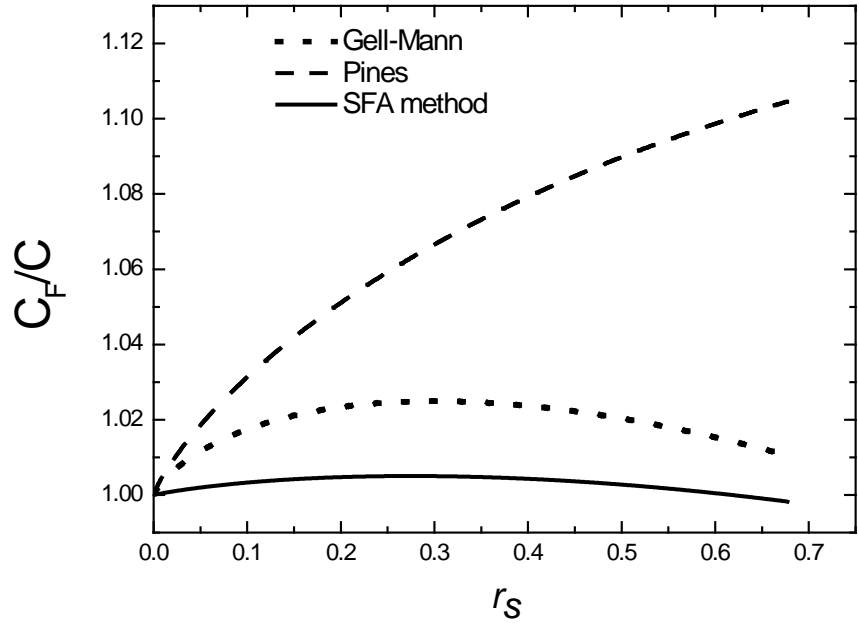

**Fig.3**. The relative expressions $C_F/C$ obtained by Gell-Mann and Pines with comparison of a similar expression (47) obtained in the frame of SFA. As one can notice from comparison of these graphs these deviations are noticeable. In the frame of the SFA method we have the minimal deviations for all range of $r_s$ comparing the result with ideal Fermi-gas.

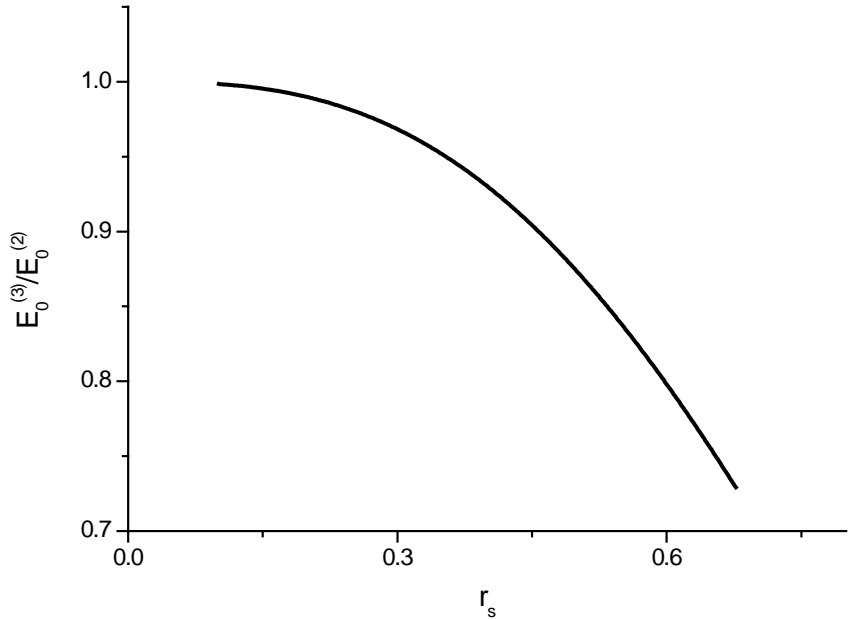

**Fig.4**. The influence of the third order corrections figuring in expression (48) with respect to expression (46). As one can notice from this comparison the contributions of the third order terms are becoming essential with increasing of the parameter $r_s$.

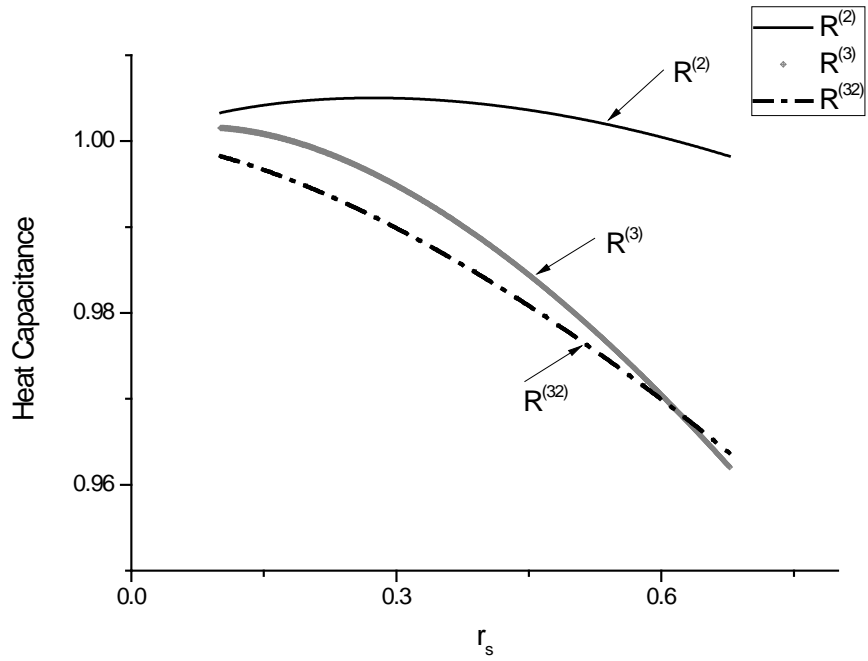

**Fig.5**. On this plot we show the comparison of the third order terms $R^{(3)}$ figuring in expression (49) with expression (47) defined here as $R^{(2)}$. As before we see that the contributions of the third order terms are becoming essential with increasing of the parameter $r_s$. The relative contributions $R^{(32)}$ of the third order terms with respect to the second order terms are shown by dash dot curve.